\documentclass{svjour3}                     
\smartqed  
\newcommand{\vlowk}{V_{{\rm low}\,k}}
\usepackage{graphicx}
\usepackage{amssymb}
\usepackage{amsmath}
\journalname{Few-Body Systems (FB20)}
\begin{document}

\title{
Electromagnetic Observables in Few-Nucleon Systems 
\thanks{Presented at the 20th International IUPAP Conference on Few-Body Problems in Physics, 20 - 25 August, 2012, Fukuoka, Japan}
}

\author{Sonia Bacca}

\institute{S. Bacca \at
              TRIUMF \\
              4004 Wesbrook Mall\\
              Vancouver, BC V6T 2A3, Canada\\
              Tel.: +1 604-222-7372\\
              Fax: +1 604-222-1074\\
              \email{bacca@triumf.ca}           %
}

\date{Received: date / Accepted: date}

\maketitle

\begin{abstract}
The electromagnetic probe is a very valuable tool to study the dynamics
of few nucleons. It can be very helpful in shedding light on the not yet fully
understood three-nucleon forces.
We present an update on the theoretical studies of electromagnetic induced reactions, such as photo-disintegration  and electron scattering off $^4$He. We will show that they potentially represent a tool to discriminate among three-nucleon forces. Then,  we will
discuss  the charge radius and the nuclear electric polarizability of the  $^6$He halo nucleus.

\keywords{Electromagnetic probe \and Reactions \and Halo nuclei}
\end{abstract}

\section{Introduction}
\label{intro}

The interactions among nucleons are governed by quantum cromodynamics (QCD). In the low energy
regime relevant to nuclear physics, QCD is not perturbative and thus difficult
to solve. A series of models have been devised in the literature to describe nuclear forces in terms of  effective degrees of freedom,  protons and neutrons.
Different approaches have been investigated, from pure 
 phenomenology to theories based on meson exchanges and more recently to the 
 systematic approach of effective field theory \cite{EM,EHM}. 
For the nucleon-nucleon ($NN$) sector, several potentials  are available that  fit  
 $NN$ scattering data with high accuracy.
On the other hand, since the nuclear potential has an  effective nature, it is in principle a many-body operator. So one expects three-body forces ($3NF$) to be important. 
Hence, a debate is taking place regarding the role of $3NF$s and how they can be constrained.
For the determination of   realistic  three-body potentials or 
to discriminate among different models, one needs to find some  few-nucleon observables involving at least three nucleons that show sensitivity to the different nuclear Hamiltonians. One also needs to perform these studies in the framework of few-body systems, where one can use exact calculations  to test the nuclear interactions from a comparison to experiment.
An important activity in this direction has taken place in recent years, 
with  calculations of bound-state and hadronic scattering properties of light nuclei. Electromagnetic observables are complementary quantities that need to be investigated as well to shed more light on the not yet fully understood three-nucleon forces.
In this paper we will present an update on recent studies of electromagnetic observables and we will limit the discussion to the well bound $^4$He nucleus and to $^6$He as an example in halo nuclei.

The paper is structured in the following way.  In Sec.~\ref{sec:1} we will first present the main calculational techniques: most of the results shown here have been obtained
using the hyper-spherical harmonics method and the Lorentz integral transform for break-up reactions.
In Sec.~\ref{sec:2} we will talk about $^4$He and in Sec.~\ref{sec:3} about $^6$He. Finally, in Sec.~\ref{sec:4} we will draw some conclusions.

\section{Theoretical tools}
\label{sec:1}

For a given  Hamiltonian $H$ one can use the hyper-spherical harmonics (HH) expansion to solve the
Schr\"{o}dinger equation $H\left|\Psi\right>=E\left| \Psi \right> $.  This method is typically employed in
few-body systems with  3 or 4 constituents, but it  can be successfully applied 
to 6-body calculations as well \cite{EIHH}.
  The approach is translationally
invariant, being constructed starting from the Jacobi coordinates. It is also very accurate for bound states.
In the HH method, the wave-function expansion reads
\begin{eqnarray}
\label{expans}
&\left| \Psi \right>  &= \sum_{[K] n}^{K_{\rm max}, n_{\rm max}}C_{[K] n} \, R_{n}(\rho) \,
{\cal Y}_{[K]}(\Omega,s_1,...,s_A, t_1, ...,t_A ),
\end{eqnarray}
where
${\cal Y}_{[K]}$ are the HH and  $R_{n}(\rho)$ are the hyper-radial wave functions, which are expressed in terms of Laguerre polynomials.   Here, $s_i$ and $t_i$ are the spin and isospin of nucleon {\it i},
respectively. $C_{[K] n}$ is the coefficient of the expansion labeled
by a cumulative quantum number $[K]$, which  includes
the grand-angular momentum $K$, and by $n$ for the hyper-radial states.

We are not only interested in bound state properties, but in electromagnetic induced reactions 
as well. 
In this case, a fundamental ingredient is the inclusive nuclear response
function 
\begin{equation}
\label{resp}
R_{O}(\omega,q)=\int \!\!\!\!\!\!\!\sum _{f} 
\left|\left\langle \Psi_{f}| O(q)| 
\Psi _{0}\right\rangle\right|
^{2}\delta\left(E_{f}-E_{0}-\omega \right)\,, 
\end{equation}
where $\omega$ and $q$ are the energy and momentum transferred,  
$| \Psi_{0/f} \rangle$ and 
$E_{0/f}$ denote initial and final state wave functions and energies. Finally ${O}$
represents a general excitation operator.
From Eq.~(\ref{resp}) it is evident that in principle one needs the knowledge of 
all  possible final states in the continuum.
This fact constitutes a major complication, which we 
circumvent by using the Lorentz integral transform (LIT) method \cite{LIT}.
This  reduces the problem to the solution of
the equation 
\begin{equation}
({H}-E_{0}-\sigma)|\widetilde{\Psi}_{\sigma,q}^{O}
\rangle={O}(q)|{\Psi_{0}}\rangle\,,\label{liteq}
\end{equation}
where $\sigma$  is a complex parameter and $|\widetilde{\Psi}_{\sigma,q}^{O} \rangle$
is a state with bound-state-like asymptotics. Eq.(\ref{liteq}) is then solved with the HH expansion.

\section{The stable $^4$He nucleus}
\label{sec:2}

In the following we will present results for electromagnetic induced reactions for the 
nucleus of $^4$He.
\begin{figure}
\includegraphics[scale=0.32,clip=]{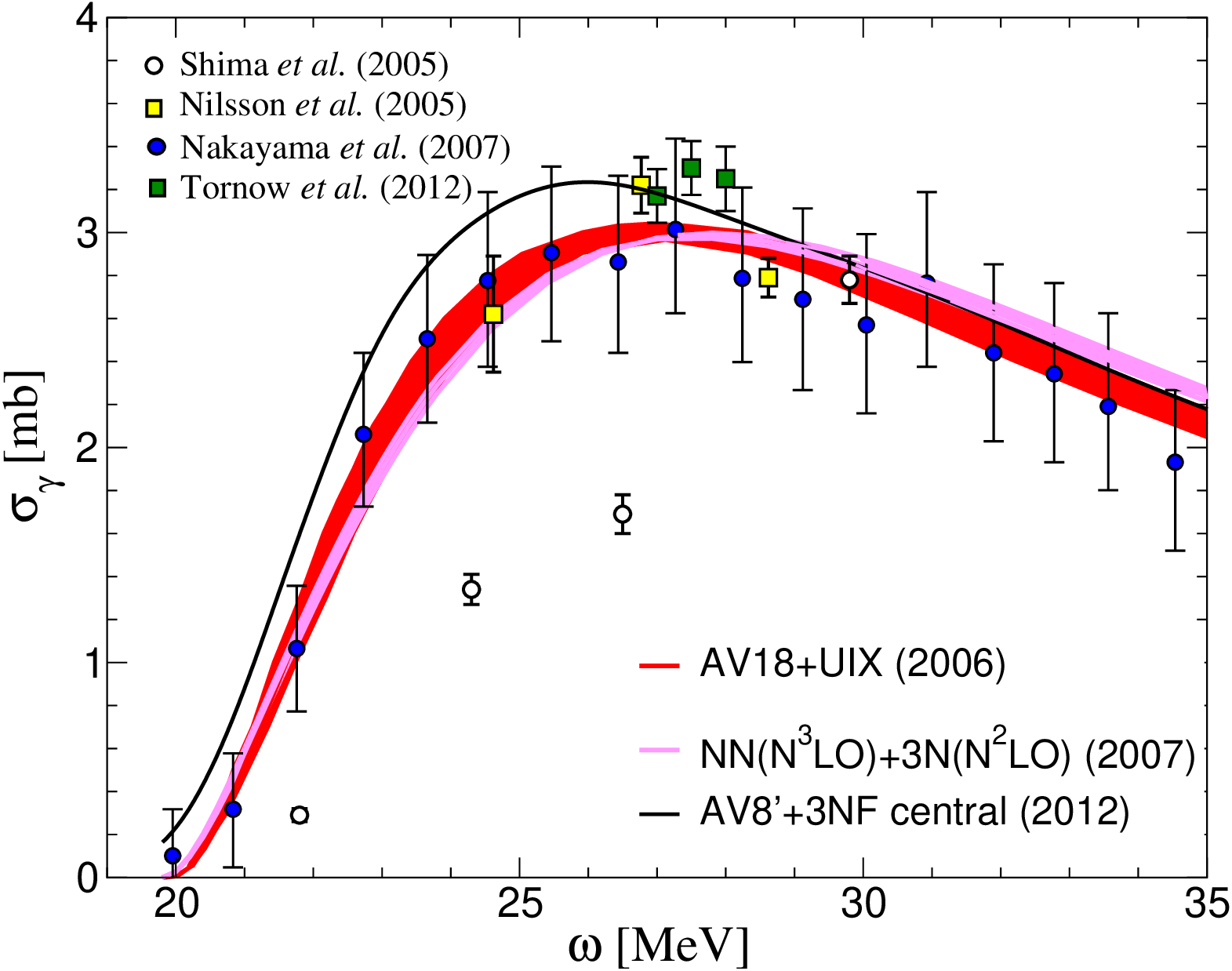}
\caption{(Color online) Total $^4$He photo-absorption cross section: 
calculations with the AV18+UIX from \cite{PRL_photon}, with EFT forces from \cite{SofiaPhoton}
and with a central $3NF$ from \cite{Wataru} in comparison with data obtained with tagged photons
(see Ref.~\cite{PRL_photon} for all experimental reference).
}
\label{fig_photon}
\end{figure}
\paragraph{Photo-disintegration cross section.}
We start our discussion from the photo-disintegration cross section of $^4$He.
 Such nucleus is the 
ideal testing ground for microscopic three-body forces,  which are 
often fitted in  three-body bound systems. Furthermore,
because of gauge invariance, nuclear forces also  manifest 
themselves  as exchange currents in photo-reactions.
In the last three decades there has been a continuous interest in the photo-disintegration 
of  $^4$He. The latest results, both from theory and  experiment, are of 2012.
The total cross section $\sigma_{\gamma}$ can be calculated as
\begin{equation}
\sigma_\gamma(\omega)=4\pi^{2}\alpha\omega R_{\rm E1}(\omega)\,,
\end{equation}
\noindent where $\alpha$ is the fine structure constant and
$R^{\rm E1}(\omega)$ is a response function as in Eq.(\ref{resp}), with
$q=\omega$. The operator is the electric dipole operator ${O}=E1$. 
The dominant part of the exchange 
currents contribution  is taken into account in the dipole response function via the 
Siegert theorem.
In Ref.~\cite{PRL_photon} we calculated $\sigma_{\gamma}$ for the first time 
using the realistic potential AV18+UIX \cite{AV18}. One year later,  $\sigma_\gamma$ 
was calculated from chiral EFT potentials using the LIT method in conjunction with the 
no core shell model \cite{SofiaPhoton}, leading to a very similar result. In 2012 a new calculation was performed with the complex scaling method \cite{Wataru} using a very simple purely central
three-body force. All these theoretical curves are shown in Fig.~\ref{fig_photon}.
 Except from threshold, where \cite{Wataru} is not accurate, the $H$-dependence of the cross section at the peak  is of the order of $10\%$.
This is much less than the difference in the experimental data.
In particular, the data from Shima {\it et~al.} \cite{Shima} are a factor of 2 smaller than all other measurements including the latest experiment from 2012 \cite{TUNL}. Because the theoretical sensitivity to changes in the Hamiltonians is smaller than the difference in the experiments, it is unfortunately not yet possible to discriminate among $3NF$s.

\paragraph{Electron scattering reaction.}
In the one-photon-exchange approximation, the inclusive cross section
for electron scattering off a nucleus is given in terms of two
response functions, i.e.  
\begin{equation}
\frac{d^2 \sigma}{d\Omega d{\omega}}=\sigma_M\left[\frac{Q^4}{q^4}
{R_L(\omega,q)}+\left(\frac{Q^2}{2 q^2}+\tan^2{\frac{\theta}{2}}\right)
{R_T(\omega, q )}\right]
\end{equation}
where $\sigma_M$ denotes the Mott cross section; $Q^2=-q_{\mu}^2={
q}^2-\omega^2$ is the squared four momentum transfer with 
$q$ being the  three-momentum transfer; 
$\theta$ is the electron scattering angle. $R_L(\omega,{q})$ and $R_T(\omega,{ q})$ are
the longitudinal 
and transverse response functions, respectively.
In the following we will discuss $R_L$, which is a function
like that in Eq.~(\ref{resp}) with $O=\rho(q)$, where $\rho({q})$ is the Fourier transform of the charge 
density operator. $R_L$ is well known for not being very sensitive
to exchange currents, so one can concentrate in studying the sensitivity of this observable to $3NF$.
\begin{figure}
\includegraphics[scale=0.22,clip=]{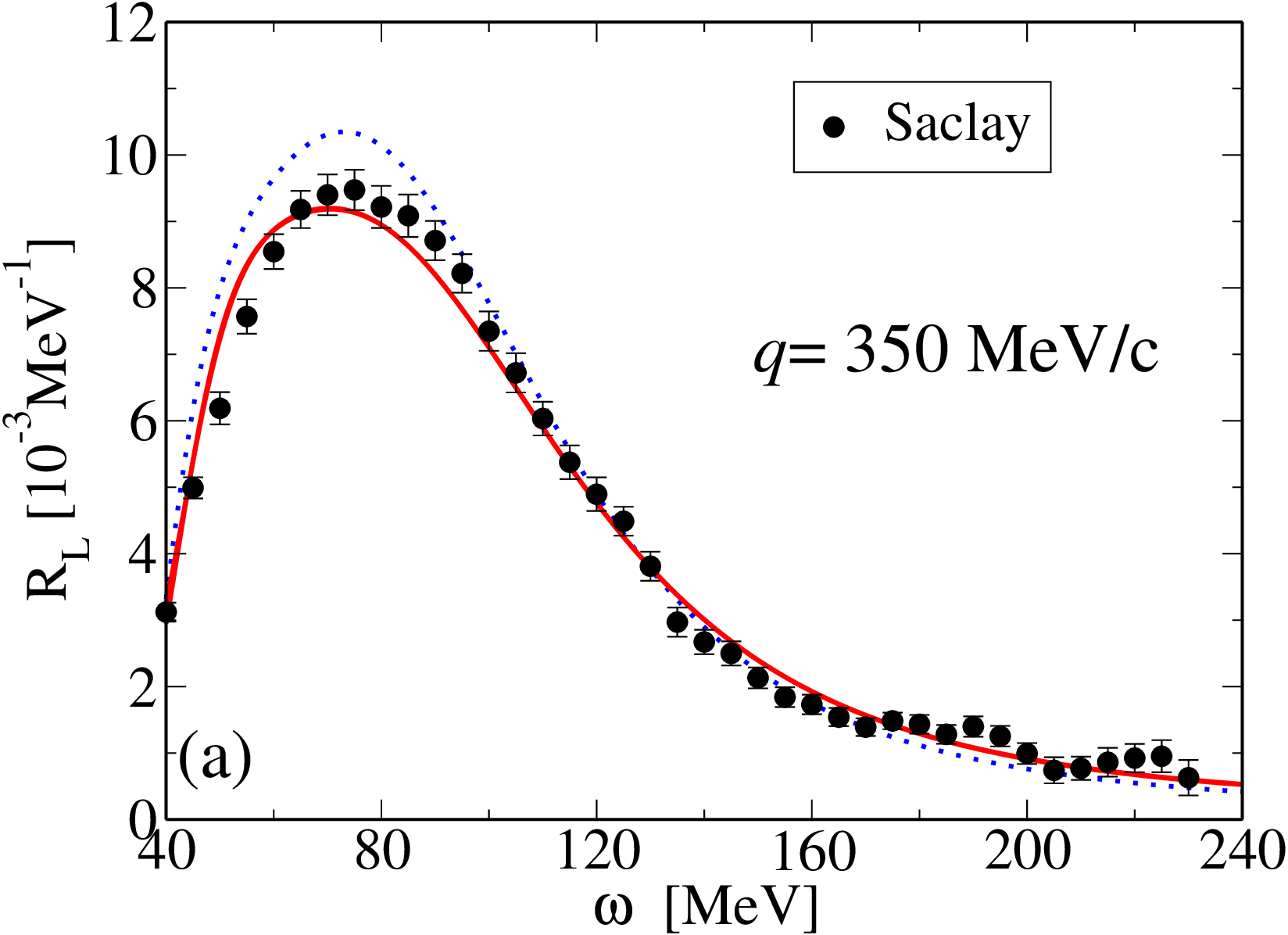}
\includegraphics[scale=0.22,clip=]{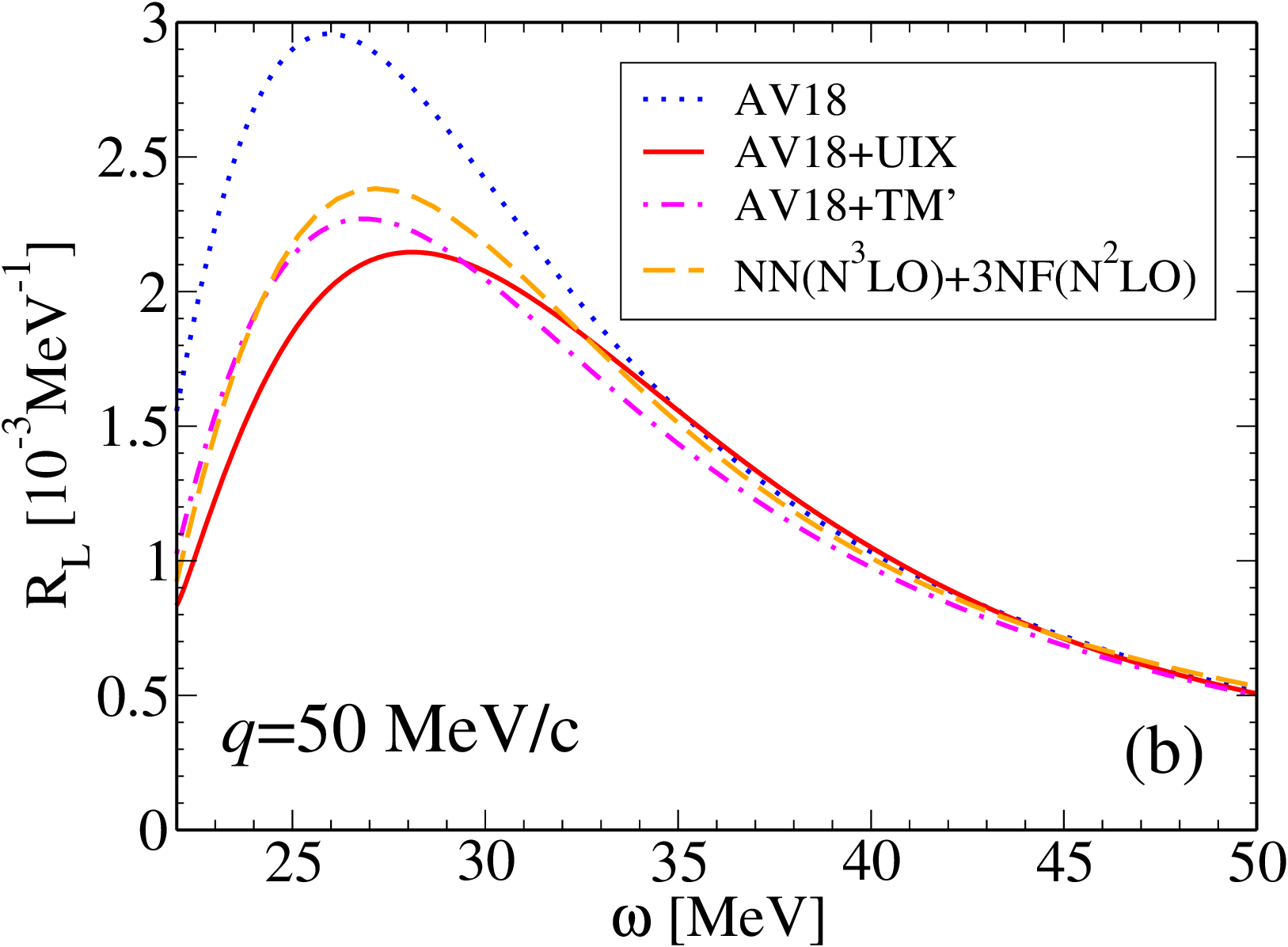}
\caption{(Color online) Panel (a): $R_L(\omega)$ at  $q=350$ MeV/c with the  AV18 (dotted),
 AV18+UIX (solid) potentials in comparison with the experimental data from Saclay.  Panel (b):
$R_L(\omega)$ at  $q=50$ MeV/c with the same potentials, with the AV18+TM' force (dash-dotted)
and a {\it preliminary} results from chiral forces (dashed) with $NN$ at N$^3$LO and $3NF$ at N$^2$LO.
} 
\label{fig_el}
\end{figure}
In Refs.~\cite{el_PRL,el_PRC} we performed for the first time a calculation of $R_L$ at different $q$ values with different realistic potentials. We found that the  experimental results are better described by theory if one includes $3NF$, as one can see from Fig.~\ref{fig_el}(a). We also performed calculations at low momentum transfer, where no experimental data have been published yet. We observed that the difference between calculations with $NN$ only and calculations which include $3NF$s is increasing at low $q$, reaching even $50\%$ at $q=50$ MeV/c.  In Fig.~\ref{fig_el}(b) we also show the sensitivity to different three-body Hamiltonians. It is such, that precise data can potentially discriminate among realistic potentials.
\begin{figure}
\includegraphics[scale=0.33,clip=]{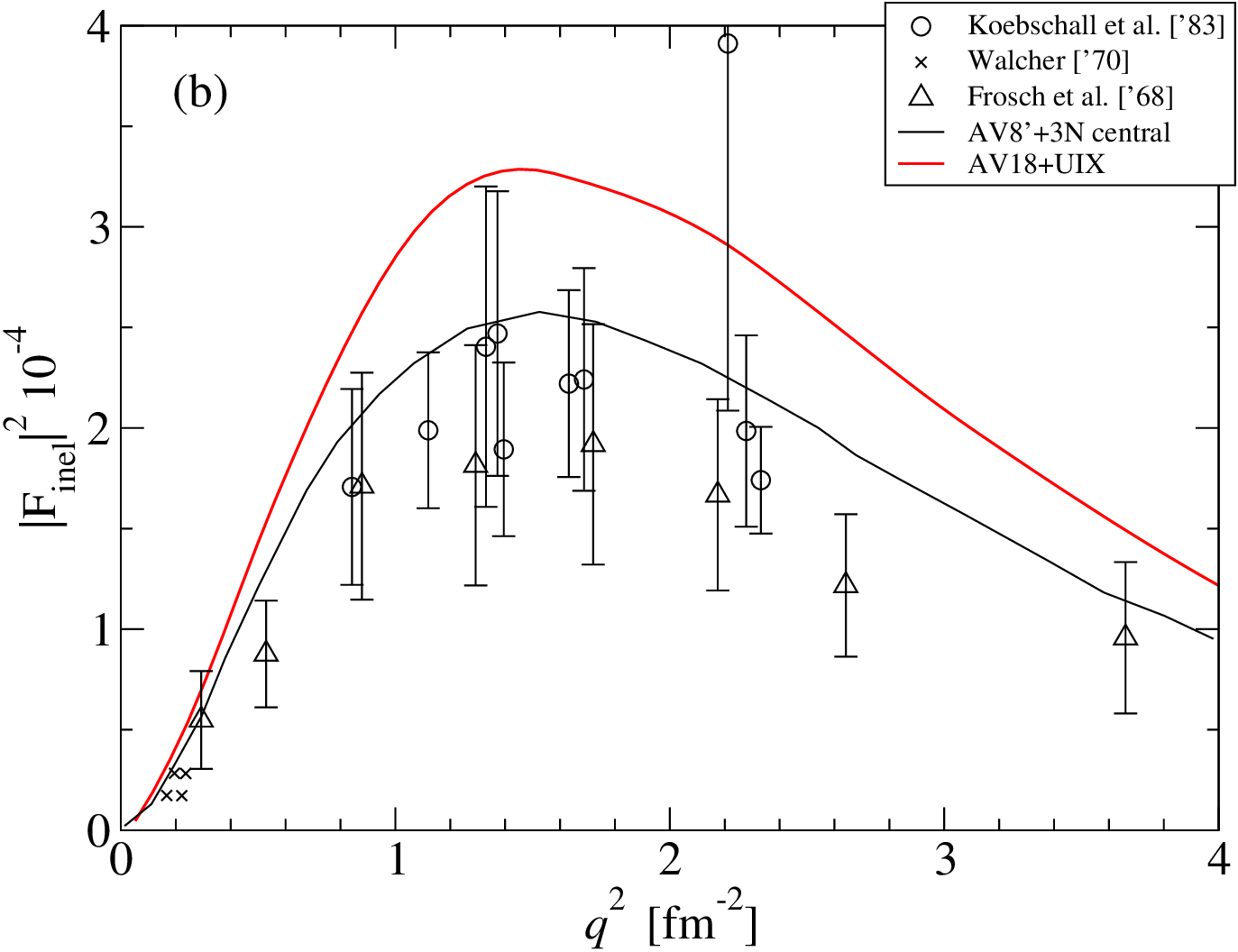}
\caption{(Color online) Inelastic form factor to the first $0^{+}$ excited state of $^4$He:
calculation with a central $3NF$ by Hiyama {\it et al.}~\cite{Hiyama} and with the AV18+UIX force ({\it preliminary}) in comparison to the available experimental data.  
\label{fig_finel}}
\end{figure}

Recently, we have studied  the inelastic transition form factor $F_{\rm inel}(q)$ to the first excited state  $0^{+}$ of $^4$He. This quantity can be also measured from electron scattering experiments and several data sets are available. Theoretically,  $F_{\rm inel}(q)$ is an inelastic observable with transitions into the continuum which are induced only by the $\ell=0$ component of $R_L$. The first calculation was performed by Hiyama {\it et al.} \cite{Hiyama} with a simple purely central $3NF$ and using bound state techniques. A good description of data was achieved. We performed a calculation of $F_{\rm inel}(q)$ treating the continuum problem with the Lorentz integral transform method and using more sophisticated $3NFs$.
Our {\it preliminary} results are shown in Fig.~\ref{fig_finel} (see other FB20 proceedings \cite{giusi_proc} for more details).
These findings point toward a large sensitivity of  $F_{\rm inel}(q)$ to the nuclear Hamiltonian, which can be potentially used to discriminate among different $3NFs$ if precise data become available.

\section{The unstable $^6$He nucleus}
\label{sec:3} 
In the following we will discuss the charge radius and the nuclear electric polarizability of $^6$He.
\paragraph{The nuclear charge radius.} 
\begin{figure}
\includegraphics[scale=0.33,clip=]{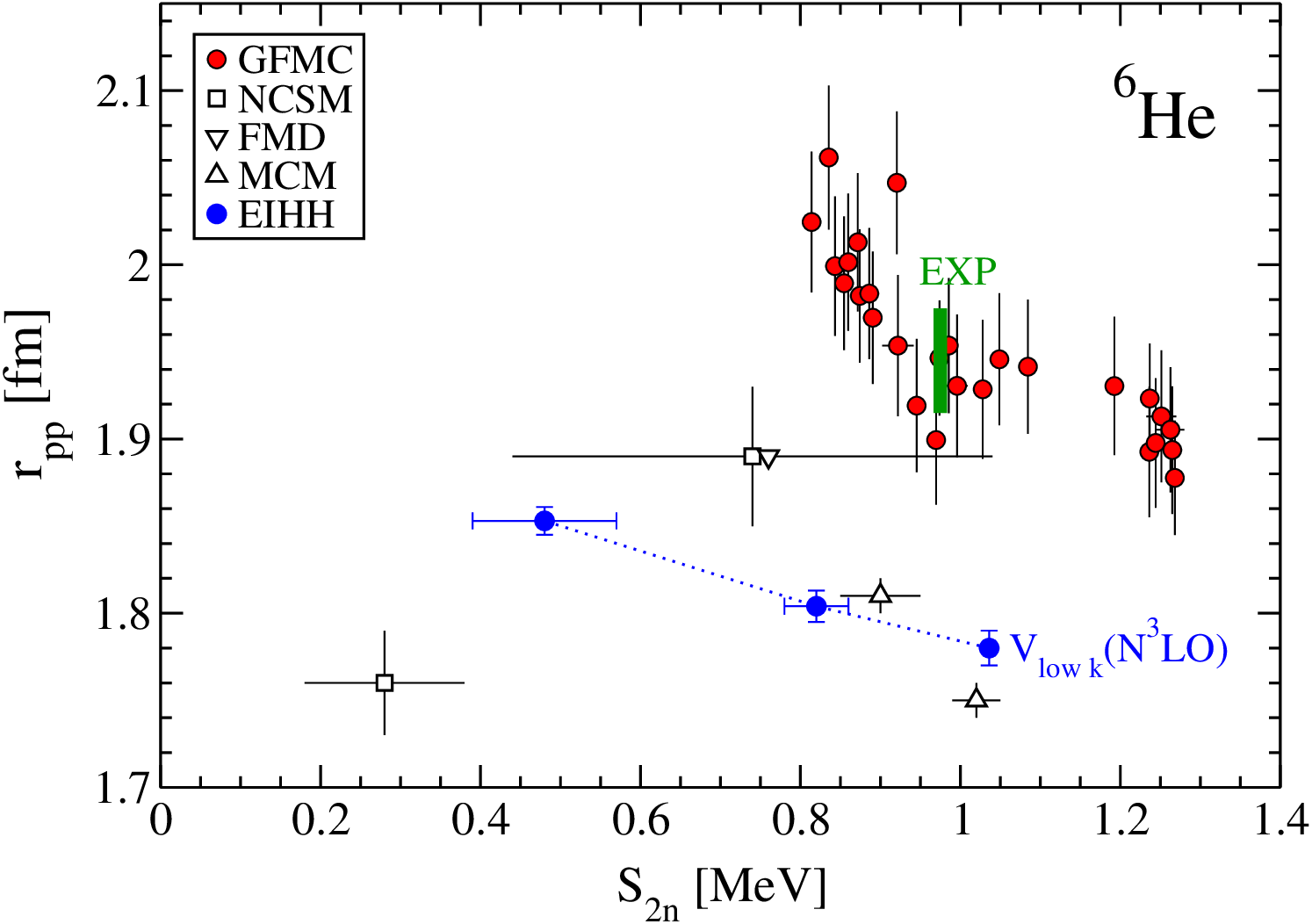}
\caption{(Color online) Correlation plot of the $^6$He  point-proton radius  versus two-neutron separation
  energy $S_{2n}$. The experimental range, shown by the bar, is compared to theory based on different methods (NCSM~\cite{NCSMHe}, FMD~\cite{FMD},
  MCM~\cite{Brida}) and to our results  with $\vlowk$. Only the
   GFMC~\cite{GFMC_Enrico_Fermi_School} calculations include $3NF$. 
} 
\label{fig:He6_correlation}
\end{figure}
  $^6$He is a
radioactive halo nucleus that undergoes $\beta$-decay with a half-life of
0.8 s. Because $^6$He is a relatively light nucleus, one can perform {\it ab-initio}
calculations and test nuclear forces from a comparison with experiment. 
Electron scattering cannot be easily performed  on  unstable nuclei, so
 to learn about the proton distribution, a viable way is to use laser spectroscopy techniques.
The charge radius $r_{\rm ch}$  of $^6$He was 
 measured  in~\cite{Wan04} and has been
recently reevaluated using input from the first direct mass
measurement~\cite{PRL_he6}, leading to $r_{\rm
  ch}=2.060\pm0.008$ fm. This can be converted into a point-proton radius, which is what {\it ab-initio} theories can calculate.
We recently have performed
 a chiral EFT based study of $^6$He~\cite{PRC_he6}, limited
to two-body forces, 
where low-momentum interactions~$\vlowk$ \cite{Vlowk} were employed
in conjunction with HH expansions of the wave function.
Here, we  present a combined comparison of our results to experiment and other {\it ab-initio} calculations (GFMC from \cite{GFMC_Enrico_Fermi_School}, NCSM from \cite{NCSMHe}, FMD from~\cite{FMD} and MCM from \cite{Brida}) by showing a plot of the  point-proton
$r_{\rm pp}$ radius versus the two-neutron separation energy $S_{2n}$ in Fig.~\ref{fig:He6_correlation}.
The cutoff $\Lambda$ dependence of our  results with $\vlowk$ 
 allows us to study the correlation between these
observables: the radius increases as the separation energy decreases,
as one expects.
Our calculations do not reproduce simultaneously $r_{\rm pp}$ and $S_{2n}$:
there exists an optimal value of $\Lambda$ where $S_{2n}$ is predicted in accordance with experiment, but $r_{\rm pp}$ is not reproduced and vice-versa. 
Also, we would like to note that 
all other calculations omit $3NF$, except from the GFMC points~\cite{GFMC_Enrico_Fermi_School}, 
which are the only ones going though the experimental band. This points towards the dependence of the
 charge radius upon $H$ and  towards the importance of including three nucleon forces.

\paragraph{The nuclear electric polarizability} 
The nuclear electric polarizability $\alpha_E$  is related to the response of the nucleus to an externally applied electric field and is relevant
in the extraction of nuclear quantities from atomic spectroscopic
measurements.  The atomic energy levels, in fact, are affected by  polarization of the
nucleus due to the electric field of the surrounding electrons.
 The polarizability of a soft halo nucleus is measured to be a lot bigger than
that of  $^4$He. Thus, it is interesting to study if this is reproduced 
by theory and whether the experimental data is explained.
The nuclear electric polarizability in the dipole approximation 
is defined by
\begin{equation}
\alpha_E=2 \alpha \sum_{f\ne0} \frac{|\langle \Psi_f| E1| \Psi_0\rangle|^2}{E_f -E_0}\,.
\label{pol}
\end{equation}
It is clearly related to the photo-absorption cross section 
by
\begin{eqnarray}
\alpha_E&=& \frac{m_{-2}(\infty)}{2\pi^2}
~{\rm with}~
m_n(\bar\omega) \,\,\, \equiv \int_{\omega_{th}}^{\bar\omega}\,d\omega \,\omega^n\,
\sigma_\gamma(\omega)\,,
\label{alphaEPSR}
\end{eqnarray}
where  $\sigma_{\gamma}$  can be calculated  exactly with the Lorentz Integral transform method. 
We recently have performed such calculation for $^6$He~\cite{he6_pol}, using a simple semi-realistic potential,
the Minnesota force, which reproduces the experimental value of the polarizability of $^4$He reasonably well. Within this force model
 one can add attractive  $P-$wave interactions  by changing the parameter $u$ (see \cite{he6_pol} for details). 
This mostly affects the binding of  $^6$He, without substantially changing $^4$He.
\begin{figure}
\includegraphics[scale=0.24,clip=]{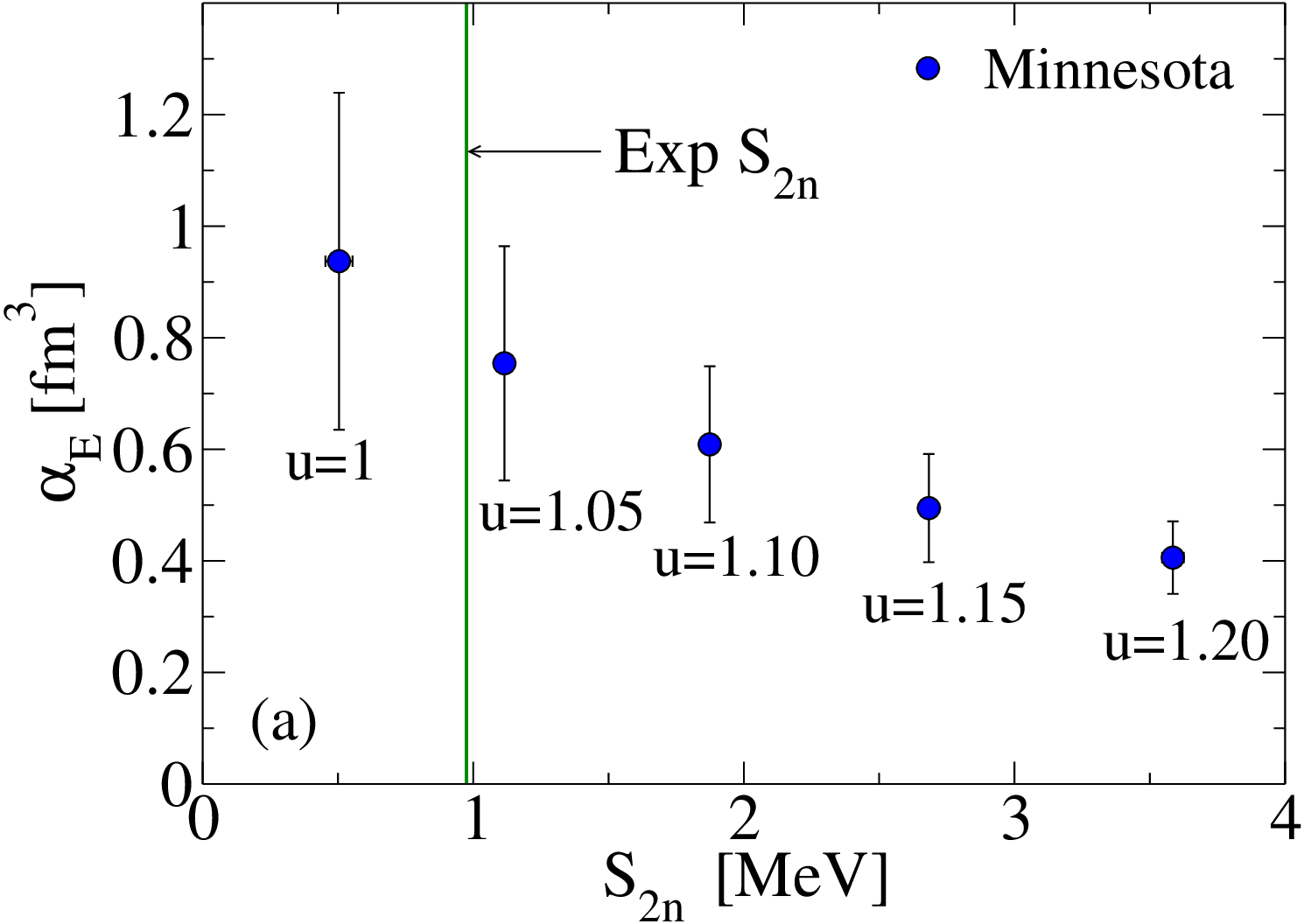}
\includegraphics[scale=0.24,clip=]{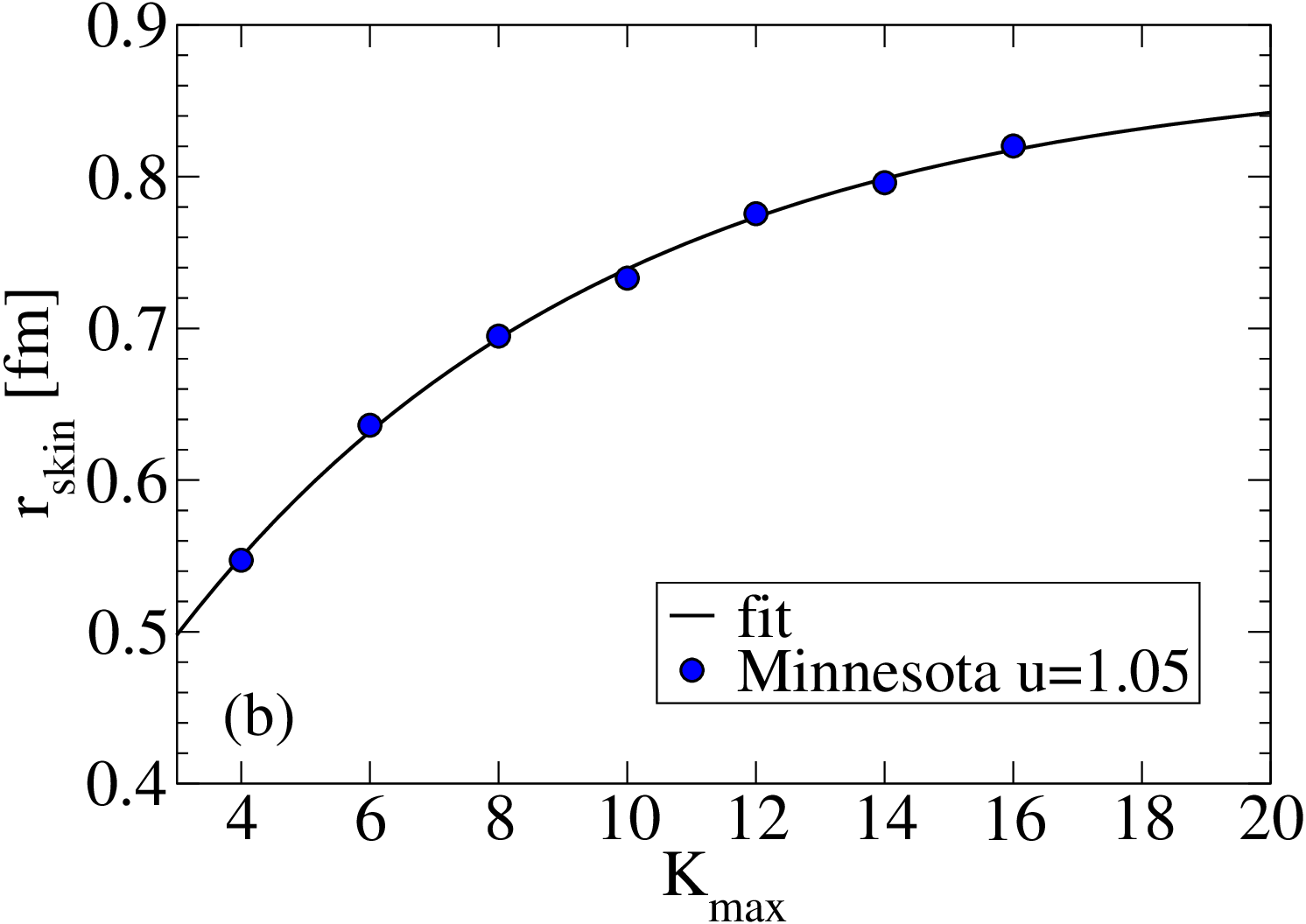}
\caption{(Color online) Panel (a): Correlation between $\alpha_E$ and
 $S_{2n}$ in $^6$He obtained with
the Minnesota potential varying the parameter $u$. The model space used here is $K_{\rm max}=12/13$.
Panel (b): Neutron skin radius $r_{\rm skin}$  of $^6$He with the Minnesota potential and $u=1.05$, as a function of the grand-angular momentum quantum number $K_{\rm max}$. 
The curve is a fit to the calculated points, used to extrapolate to infinite model space.
} 
\label{fig_pol}
\end{figure}
By varying $u$ we first observed a correlation of $\alpha_E$ vs $S_{2n}$, as shown in Fig.~\ref{fig_pol}(a).
We have chosen $u$ so that the halo feature, represented by $S_{2n}$, is reproduced.
We have then studied the correlations between other two observables:   $\alpha_E$ and the skin radius $r_{\rm skin}$. The latter is defined as
$r_{\rm skin}=r_n -r_p$, 
where $r_n$ and $r_p$ are the mean point-neutron and point-proton radii.
By varying the HH model space we observed that 
 $\alpha_E$ and $r_{\rm skin}$ are correlated linearly as
$\alpha_E=a+b~r_{\rm skin}$.
From our theoretical data we have
 ``measured'' the coefficients $a$ and $b$
and then we used them to estimate the polarizability
out of a bound-state calculation.
The calculation of $r_{\rm skin}$, in fact, does not require an expansion on the dipole excited states and as such is less computationally demanding and can be performed for larger model spaces ($K_{\rm max}=16$). The convergence of $r_{\rm skin}$ is shown in Fig.~\ref{fig_pol}(b).
Extrapolating the points with an exponential {\it ansatz} of the form~ $r_{\rm skin}(K_{\rm max})=r_{\rm skin}({\infty})-c e^{-\kappa K_{\rm max}}$
we get $r_{\rm skin}(\infty)=0.87(5)$ fm.
Using this skin radius and the linear dependence,  we
estimate the theoretical nuclear electric polarizability of $^6$He to be  
$\alpha_E=1.00(13)$ fm$^3$. Even though theory correctly predicts $\alpha_E(^6$He)
 to be one order of magnitude larger than the $\alpha_E(^4$He),  $\alpha_E(^6$He) is about a factor of two smaller than the experimental value \cite{Moro} of $\alpha_E^{\rm exp}=1.99(40)$ fm$^3$. This points toward a potential disagreement of theory with experiment. The sensitivity of $\alpha_E$ with
respect to different $H$, that reproduce $S_{2n}$ will be investigated in the future.
\section{Conclusions}
\label{sec:4}
In conclusion, because electromagnetic observables  show sensitivity to the nuclear Hamiltonians, they need to be further investigated both in theory and experiment. They have the potential of shedding more light on the not yet fully understood three-nucleon forces.

\begin{acknowledgements}
 It is a pleasure to thank the organizers  for the very interesting conference and for providing a stimulating environment.
I would like to thank my collaborators Nir Barnea, Raymond Goerke, Winfried Leidemann, Giuseppina Orlandini and Achim Schwenk, whose help has been fundamental in obtaining the results shown in this proceedings.
This work was supported  by the
Natural Sciences and Engineering Research Council (NSERC) and by the
National Research Council of Canada.
\end{acknowledgements}



\end{document}